\documentclass[3p,times,procedia]{elsarticle}
\usepackage{nupha_ecrc}
\usepackage{epsfig}
\usepackage{natbib}
\usepackage{epstopdf}
\usepackage{mathrsfs}
\usepackage{slashed}
\usepackage{amsmath}
\usepackage{verbatim}
\usepackage{graphicx}
\usepackage{amssymb}
\usepackage{psfrag}
\usepackage{array}
\usepackage{lipsum}
\usepackage{subcaption}

\volume{00}

\firstpage{1}

\journalname{Nuclear Physics A}

\runauth{}

\jid{nupha}

\jnltitlelogo{Nuclear Physics A}

\usepackage{amssymb}

\usepackage[figuresright]{rotating}

\begin{document}

\begin{frontmatter}



\dochead{XXVIIIth International Conference on Ultrarelativistic Nucleus-Nucleus Collisions\\ (Quark Matter 2019)}

\title{Identifying the nature of the QCD transition in heavy-ion collisions with deep learning}


\author[1,2,3,4]{Yi-Lun Du}

\author[1]{Kai Zhou}

\author[1]{Jan Steinheimer}

\author[5,6,7]{Long-Gang Pang}

\author[1,2]{Anton Motornenko}

\author[3,8,9]{Hong-Shi Zong}

\author[5,6,7]{Xin-Nian Wang}

\author[1,2,10]{Horst St\"{o}cker}

\address[1]{Frankfurt Institute for Advanced Studies, Giersch Science Center, D-60438 Frankfurt am Main, Germany}
\address[2]{Institut f$\ddot{u}$r Theoretische Physik, Goethe Universit$\ddot{a}$t Frankfurt, D-60438 Frankfurt am Main, Germany}
\address[3]{Department of Physics, Nanjing University, Nanjing 210093, China}
\address[4]{Department of Physics and Technology, University of Bergen, 5007 Bergen, Norway}
\address[5]{Nuclear Science Division, Lawrence Berkeley National Laboratory, Berkeley, California 94720, USA}
\address[6]{Physics Department, University of California, Berkeley, CA 94720, USA}
\address[7]{Key Laboratory of Quark and Lepton Physics (MOE) and Institute of Particle Physics, CCNU, Wuhan 430079, China}
\address[8]{Nanjing Proton Source Research and Design Center, Nanjing 210093, China}
\address[9]{Department of physics, Anhui Normal University, Wuhu, Anhui 241000, China}
\address[10]{GSI Helmholtzzentrum f$\ddot{u}$r Schwerionenforschung, D-64291 Darmstadt, Germany}

\begin{abstract}
In this proceeding, we review our recent work using deep convolutional neural network (CNN) to identify the nature of the QCD transition in a hybrid modeling of heavy-ion collisions. Within this hybrid model, a viscous hydrodynamic model is coupled with a hadronic cascade ``after-burner". As a binary classification setup, we employ two different types of equations of state (EoS) of the hot medium in the hydrodynamic evolution. The resulting final-state pion spectra in the transverse momentum and azimuthal angle plane are fed to the neural network as the input data in order to distinguish different EoS. To probe the effects of the fluctuations in the event-by-event spectra, we explore different scenarios for the input data and make a comparison in a systematic way. We observe a clear hierarchy in the predictive power when the network is fed with the event-by-event, cascade-coarse-grained and event-fine-averaged spectra. The carefully-trained neural network can extract high-level features from pion spectra to identify the nature of the QCD transition in a realistic simulation scenario.
\end{abstract}


\end{frontmatter}

\section{Introduction}\label{sec:intro}
A new state of matter, quark-gluon plasma (QGP), is predicted by the QCD theory if the temperature of strongly-interacting matter becomes high enough. Lattice QCD has established that the transition from a hadron gas to the QGP is a smooth crossover at a high temperature and low net baryon density. A first-order phase transition is conjectured at low temperature and moderate to high net baryon densities. The main goal of relativistic heavy ion experiments at the RHIC and LHC is to search for signals for the QCD phase transition and study the properties of QGP.
Phenomenologically, conventional method to study the properties of QGP is to compare model simulations with experimental data by varying parameter sets and employing different EoSs. Recently, modern statistical methods such as Bayesian methods are used to fit a set of different observables globally to constrain the properties of QGP~\cite{pratt2015constraining,bernhard2016applying}. However, with these methods one cannot obtain direct correlations between observables and the focused property of QGP rather than others. Machine learning methods are possible tools to tackle such tasks. Recently, a deep CNN classifier is developed to identify the nature of the QCD transition with a high predictive accuracy $\thicksim95\%$ in pion spectra from a pure hydrodynamic study~\cite{pang2018equation}, which are robust against different initial conditions, physical parameters and hydrodynamics solvers. In this proceeding, we review the generalizability of this method in a more realistic scenario of heavy ion collisions where the hadronization, hadronic rescattering and resonance decays after the hydrodynamics evolution are taken into account in a hybrid model~\cite{du2020identifying}. Due to the finite number of particles and the stochastic dynamics, the discrete event-by-event pion spectra have significant fluctuations that might blur out the correlations one is looking for.

\section{Micro-Macro hybrid model of relativistic heavy-ion collisions}\label{sec:model}
\begin{figure}[!bht]
\begin{subfigure}{0.46\textwidth}
 \centering
 \includegraphics[width=1.0\textwidth]{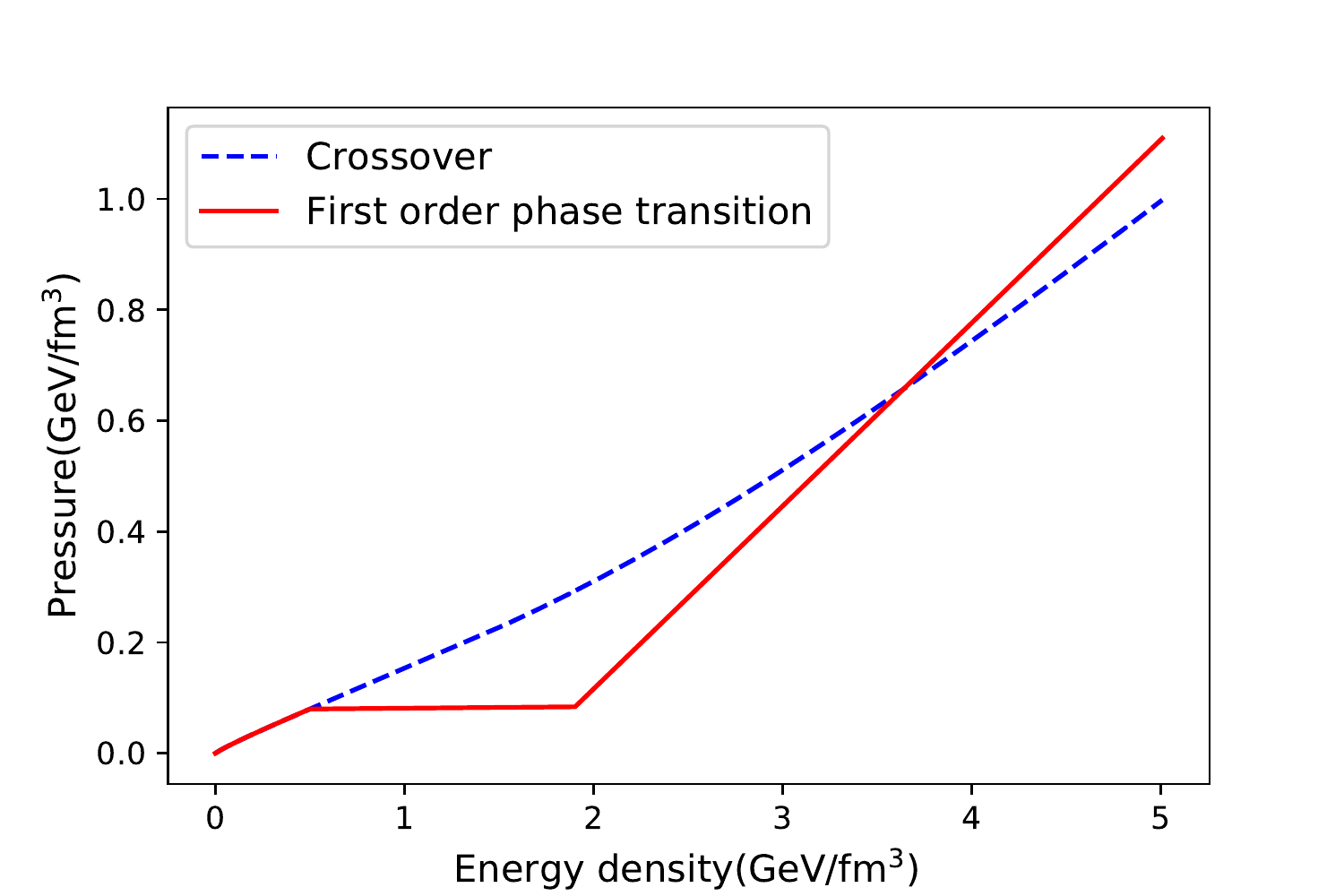}
 \caption{ }
 \label{eos}
\end{subfigure}
\begin{subfigure}{0.54\textwidth}
 \centering
 \includegraphics[width=1.0\textwidth]{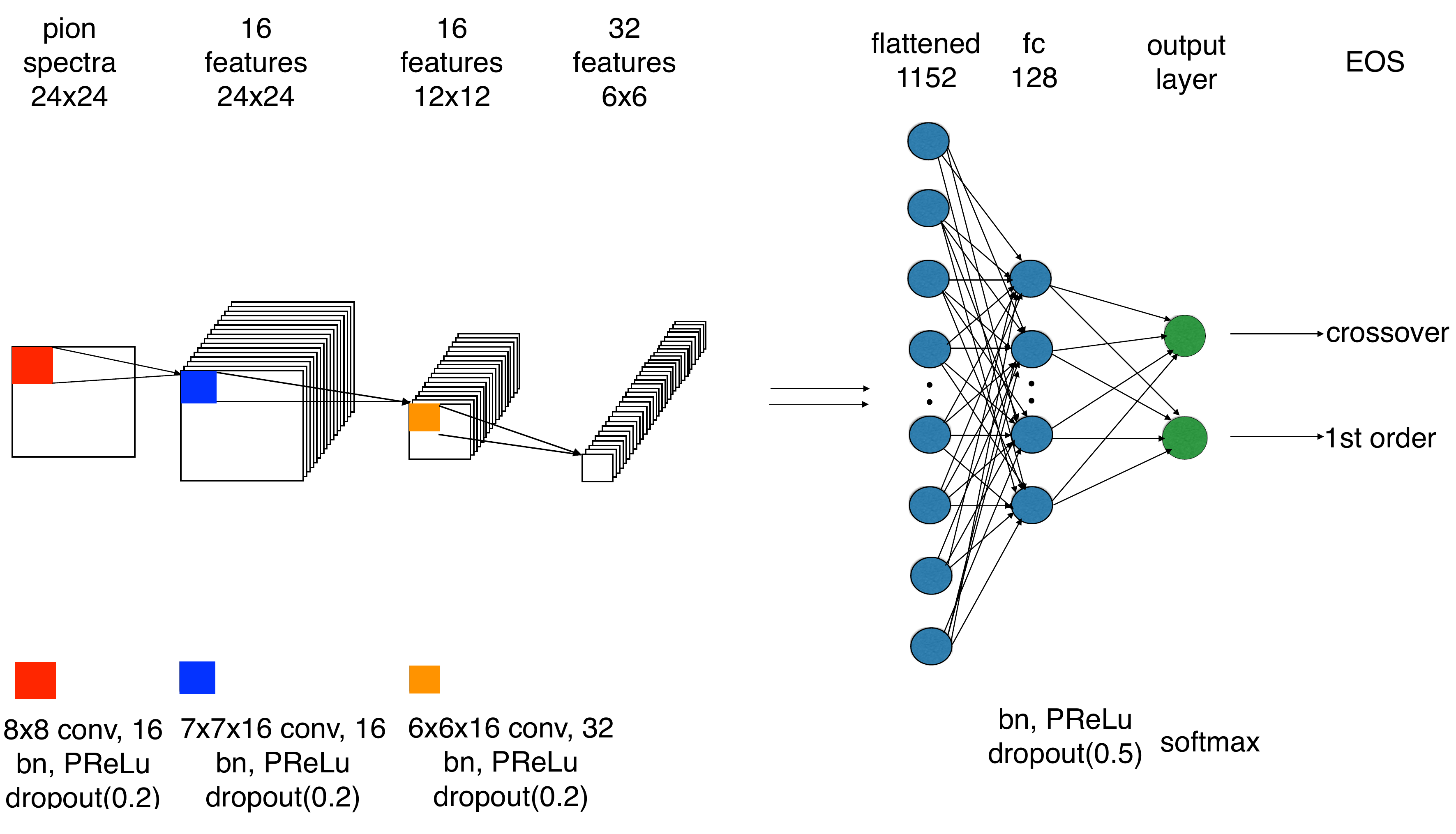} 
 \caption{ }
 \label{cnn}
\end{subfigure}
\caption{Left: Two different EoSs are implemented in the hydrodynamic simulation, pressure as functions of the energy density. A crossover one is compared with a first order phase transition with a transition temperature $T_c=165$ MeV. The baryon-chemical potential is is assumed to be exactly $\mu_B=0$ throughout the whole work. Right: The CNN architecture.}
\end{figure}

We use the iEBE-VISHNU hybrid model~\cite{shen2016iebe} to perform event-by-event simulations of relativistic heavy-ion collisions at RHIC and LHC energies. This hybrid model uses the MC-Glauber~\cite{broniowski2009glissando,alver2008phobos} or MC-KLN~\cite{kharzeev2005onset,kharzeev2005color} model to generate fluctuating initial conditions. The hydrodynamics simulation uses two different types of EoSs: (1) a crossover EoS from a lattice-QCD parametrization~\cite{huovinen2010qcd}; (2) a first-order EoS from a Maxwell construction between a hadron resonance gas and an ideal gas made of quarks and gluons with transition temperature $T_c=165$ MeV~\cite{sollfrank1997hydrodynamical}. These two EoSs are depicted in Fig.~\ref{eos}. 
After the hydrodynamic evolution, the fluid fields are projected into hadrons via the Cooper-Frye formula, and a hadronic cascade follows with the UrQMD model~\cite{bass1998microscopic,bleicher1999relativistic} where resonance decays and hadronic rescatterings are both included.

This hybrid model can fit experimental data on final hadron spectra by varying adjustable parameters which include: the equilibration time $\tau_0$, the ratio of the shear viscosity to the entropy density $\eta/s$ and the switching temperature $T_{sw}$ from hydrodynamics to hadronic cascade stage. We vary these model parameters in the generation of the training data to encourage the neural network to capture the intrinsic features encoded in the EoS, rather than those biased by the specific setup of other uncertain physical properties. 

\section{Neural network and input data}\label{sec:neural network}
The CNN architecture used here is shown in Fig.~\ref{cnn} whose details can be found in Ref.~\cite{du2020identifying}. The input to this neural network is a histogram of the number of pions at mid-rapidity $|y|\leq 1$ $\rho(p_T,\Phi) \equiv dN_{\pi}/dy dp_T d\Phi$ with 24 $p_T$-bins and 24 $\Phi$-bins. $p_T$ denotes the transverse momenta, while $\Phi$ denotes the azimuthal angles.

We explore six different scenarios for the input to the neural network in a systematic way. A late- and early-transition from hydrodynamics to the hadronic cascade are both considered by taking the switching temperature the same value as the freeze-out temperature used in pure hydrodynamics study~\cite{pang2018equation}, $T_{sw}=137$ MeV and a realistic value $T_{sw}>150$ MeV, respectively. In the late-transition scenario, the duration of the hadronic cascade are significantly diminished and the effects of the finite number of particles and resonance decays are mainly left as compared to the pure hydrodynamics modeling. The early-transition scenario is more realistic and different from the late-transition one in two aspects. Firstly, the contribution from resonance decays in the pion spectra is increased. Secondly, the elongated duration of the hadronic cascade might further erase the imprint of EoS encoded in the pion spectra.

In both of the late- and early-transition scenarios, we have three sub-scenarios for the input to the neural network: {\it event-by-event spectra}, {\it cascade-coarse-grained spectra} and {\it event-fine-averaged spectra}. To mitigate the fluctuations in the event-by-event spectra due to the finite number of particles, hadronic cascade and resonance decays, certain averaging procedures are possible methods. In the model simulations, one can repeat the hadronic cascade for any times for the same hydrodynamic evolution. The cascade-coarse-grained spectra are explored, where the spectra of 30 such simulations are coarse-grained by averaging as to smear the fluctuations induced by the cascade. However, one defect of such averaging procedure is that the separation of dynamics between hydrodynamic and hadronic cascade stage is purely theoretical. Thus an experimentally controllable averaging procedure is favored. The event-fine-averaged spectra, which are the average over the spectra of 30 events generated within the same fine centrality bin of 1\% width are explored. In such spectra, the initial-state fluctuations are also significantly mitigated.

\section{Results and Conclusion}\label{sec: train-test}
\begin{figure}[!bth]
\begin{subfigure}{.5\textwidth}
 \centering
 \includegraphics[width=1.0\textwidth]{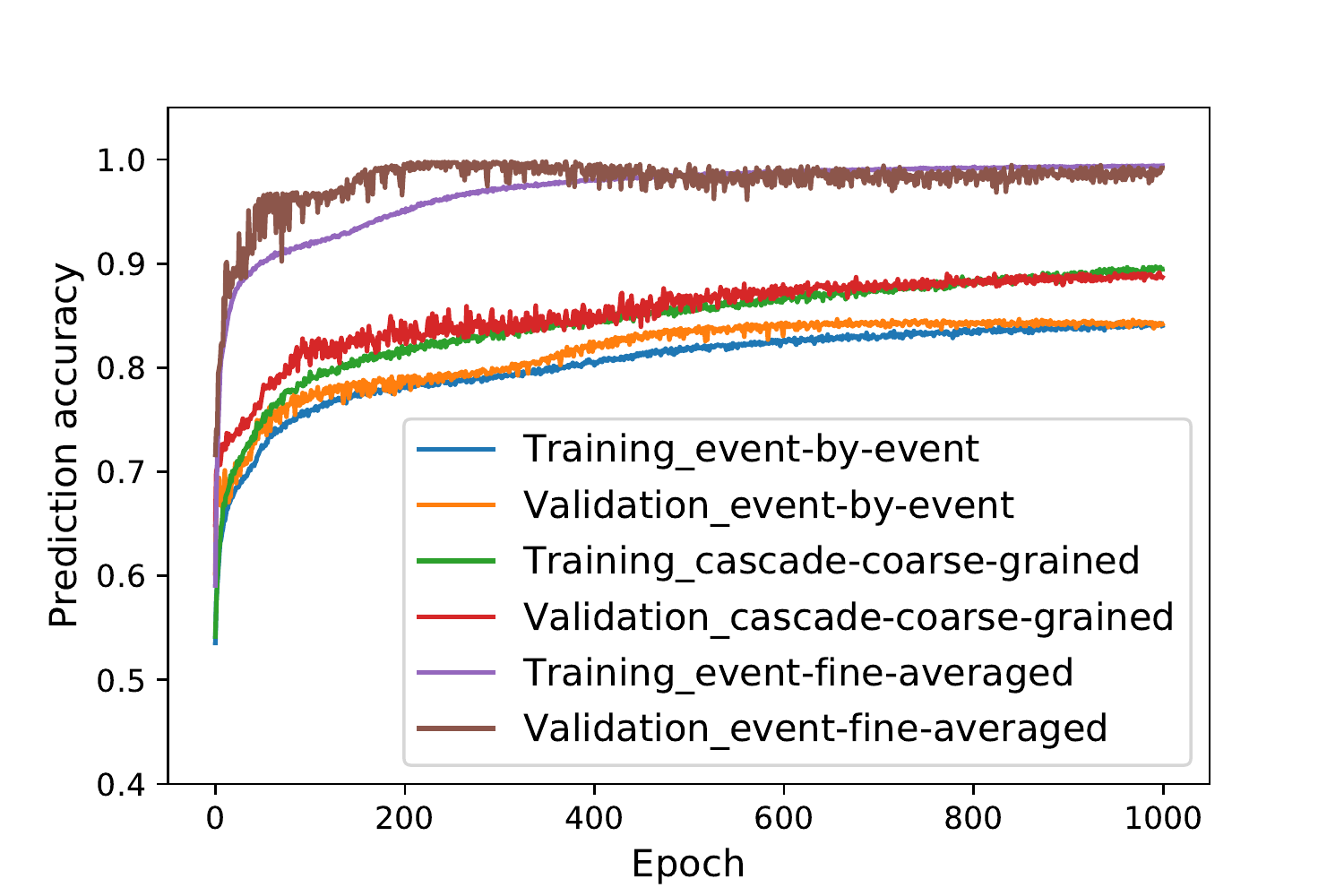} 
 \caption{ }
 \label{fig-accuracyloss}
\end{subfigure}
\begin{subfigure}{.5\textwidth}
 \centering
 \includegraphics[width=1.0\textwidth]{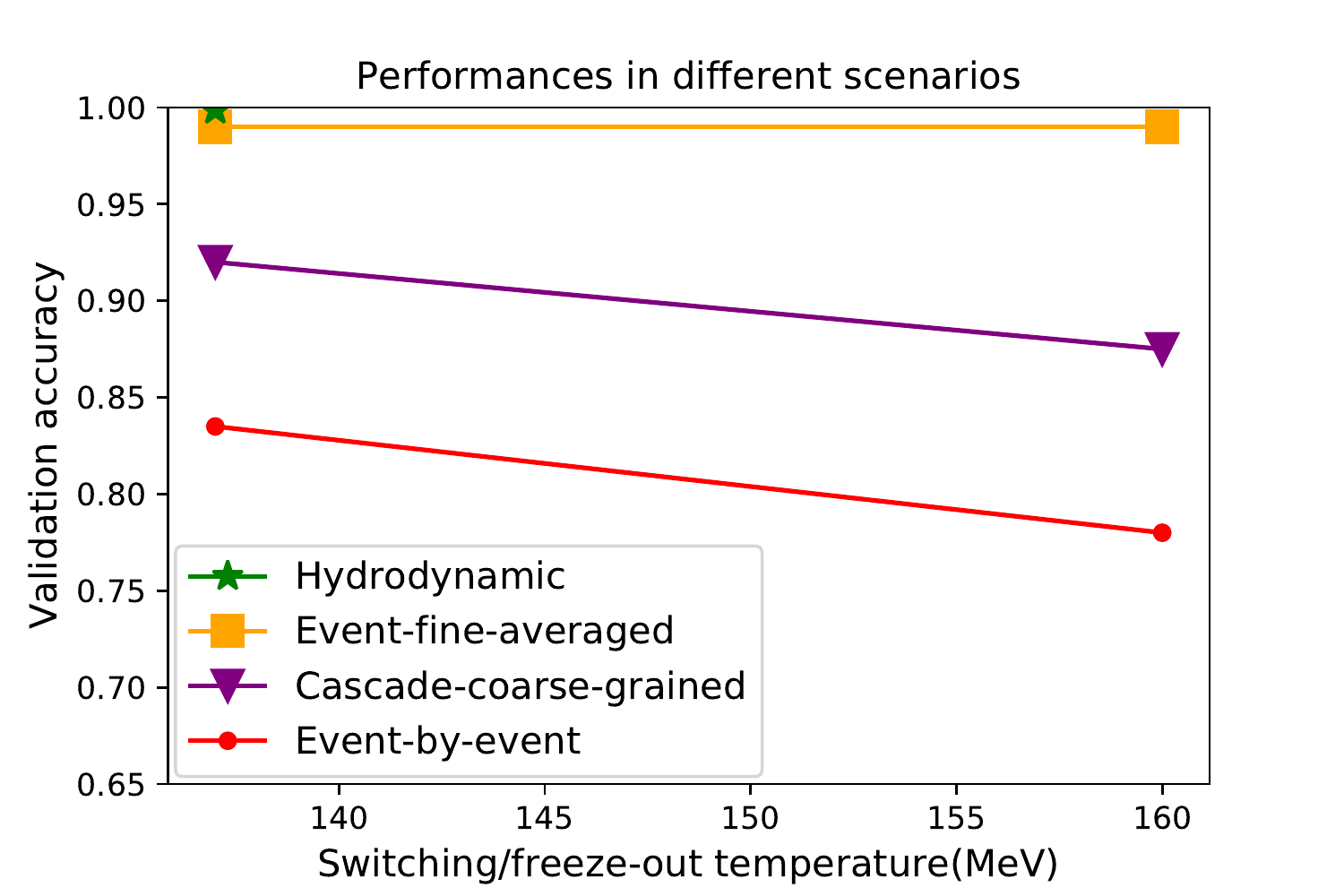} 
 \caption{ }
 \label{Fig: Performance}
\end{subfigure}
\caption{(a) Training and validation accuracy in training history for three sub-scenarios of the late-transition scenario with $T_{sw}=137$ MeV. These three sub-scenarios refer to the event-by-event spectra (blue and orange), the cascade-coarse-grained spectra (red and green) as well as the 30-events-fine-averaged spectra (purple and brown). (b) Comparison between the validation accuracy in all the scenarios studied. The green star depicts the pure hydrodynamic result~\cite{pang2018equation}. The red filled circle, the purple triangle and the orange square symbols depict the performance given by the event-by-event spectra, cascade-coarse-grained and 30-events-fine-averaged, respectively, with different switching temperatures.}
\label{fig: result}
\end{figure}
Fig.~\ref{fig-accuracyloss} shows the training history of the CNN in three aforementioned sub-scenarios in the late-transition scenario. In each sub-scenario, training and validation accuracy are still close after 1000-epochs training, which implies that over-fitting issue is avoided since the validation data are generated with the same simulation parameters as the training data but are never seen by the network in the training. A clear hierarchy of the prediction accuracy can be observed when the averaging procedure is performed over more and more stages of the whole dynamics. The lowest accuracy, about 80\%, is given by the CNN with event-by-event spectra, the moderate accuracy, about 90\%, is given by the CNN with cascade-coarse-grained spectra, while the highest accuracy, about 99\%, is given by the CNN with the 30-events-fine-averaged spectra, which is as high as in the pure hydrodynamic study~\cite{pang2018equation}.  This hierarchy shows that proper reduction of fluctuations from either the final hadronic cascade or together with the initial conditions in the pion spectra will help the CNN to reveal the EoS information. 

Fig.~\ref{Fig: Performance} summarizes the predictive performances on the validation data in all the aforementioned exploratory scenarios.  The slight downward trend for the validation accuracy of network with respect to the switching temperature, implies that the elongated resonance decays and hadronic cascade will blur out the correlation between the EoS information in the hydrodynamics and the final-state pion spectra. We mention that the testing accuracy on the data which are generated with different initial conditions and simulation parameters is generally a little lower than the validation one, which demonstrates that the performance of the trained CNN is robust against different model setups including initial conditions, $\tau_0$, $\eta/s$ and $T_{sw}$ within certain ranges.

We extended a previous work on identifying EoS from pion spectra in the pure hydrodynamics modeling of heavy ion collisions using deep learning technique. In this work, we explore a more realistic hybrid modeling for heavy-ion collisions, where hadronization, hadronic cascade and resonance decays are properly considered. It is demonstrated that, even after these stochastic dynamics, the imprint of EoS in hydrodynamics evolution survives in the final-state pion spectra, from the perspectives of deep CNN. In the future it's tempting to explore how to generalize this method to the data from other simulation models as well as experimental data.

\section*{Acknowledgement}
This work is supported by the HGS-HIRe for FAIR, by the GSI F\&E, by the AI grant of SAMSON AG, by the BMBF, and by the Judah M. Eisenberg Laureatus Chair and the Walter Greiner Gesellschaft, by Trond Mohn Foundation under Grant No. BFS2018REK01, by National Natural Science Foundation of China under Grant Nos.11475085, 11535005, 11690030 and 11221504, and National Major state Basic Research and Development of China under Grant Nos. 2016Y-FE0129300 and 2014CB845404, and the U.S. Department of Energy under Contract Nos. DE-AC02-05CH11231, and the U.S. National Science Foundation (NSF) under Grant No. ACI-1550228 (JETSCAPE).





\bibliographystyle{elsarticle-num}
\bibliography{duyl}







\end{document}